\pdfoutput=1
\documentclass[preprint,12pt]{elsarticle}
\biboptions{sort&compress}

\usepackage[utf8]{inputenc}
\usepackage[T1]{fontenc}
\usepackage{amsmath}
\usepackage{amssymb}
\usepackage{bm}

\usepackage{graphicx}
\usepackage{float}
\usepackage{booktabs}
\usepackage{multirow}
\usepackage{tikz}
\usetikzlibrary{arrows.meta,calc,positioning}
\usepackage{algorithm}
\usepackage{algpseudocode}

\usepackage{subcaption}

\usepackage{xcolor}
\usepackage{hyperref}

\hypersetup{
    colorlinks=true,
    linkcolor=blue,
    citecolor=blue,
    urlcolor=blue
}
\graphicspath{{figures/}}

\newcommand{\mumag}{$\mu$Mag}
\newcommand{\Ms}{M_{\mathrm{s}}}
\newcommand{\Aex}{A}
\newcommand{\Ku}{K_{\mathrm{u}}}
\newcommand{\Heff}{\mathbf{H}_{\mathrm{eff}}}
\newcommand{\Hdemag}{\mathbf{H}_{\mathrm{d}}}
\newcommand{\Happ}{\mathbf{H}_{\mathrm{app}}}
\newcommand{\mvec}{\mathbf{m}}
\newcommand{\Mvec}{\mathbf{M}}

\newcommand{\LLG}{Landau-Lifshitz-Gilbert}

\begin{document}

\begin{frontmatter}

\title{
A Heterogeneous CPU--GPU Framework for\\
Finite Element Micromagnetic Simulations\\
on Tetrahedral Meshes
}

\author[math]{Zekai Zheng}
\author[cuhksz-sse]{Zixuan Cui}
\author[math,skl-math]{Guanghui Hu\corref{cor1}}
\ead{garyhu@um.edu.mo}
\cortext[cor1]{Corresponding author.}

\affiliation[math]{
    organization={Department of Mathematics, University of Macau},
    city={Macao S.A.R.},
    country={China}
}

\affiliation[cuhksz-sse]{
    organization={School of Science and Engineering, The Chinese University of Hong Kong, Shenzhen},
    city={Shenzhen, Guangdong},
    country={China}
}

\affiliation[skl-math]{
    organization={State Key Laboratory of Internet of Things for Smart City, University of Macau},
    city={Macao S.A.R.},
    country={China}
}

\begin{abstract}
The design of magnetic storage materials relies on repeated micromagnetic simulations. Within each simulation, the demagnetization field evaluation and the sparse linear solve are executed at every time step and together dominate the computational cost. To address these bottlenecks, we develop a heterogeneous CPU-GPU framework that exploits the complementary computational capabilities of the two processors.  In the existing AFEMAG solver, developed by Yang et al [J. Comput. Phys. 431 (2021) 110142], computationally intensive boundary integral evaluations and sparse linear solves limit performance. We offload these operations to the GPU while retaining assembly and iteration control on the CPU. We validate the framework using cases from all five \mumag{} standard problems, covering quasistatic hysteresis, equilibrium configurations, field-driven dynamics, and spin-transfer-torque dynamics. This approach accelerates simulations across the benchmark suite; notably, for Standard Problem 3 on a mesh of 137,313 nodes, augmenting the CPU with one NVIDIA A100 GPU yields a $17.1\times$ speedup over the CPU-only configuration. The heterogeneous AFEMAG framework thus provides an efficient and practical platform for micromagnetic simulations in the design of magnetic storage materials.

\end{abstract}


\end{frontmatter}
\section{Introduction}
\label{sec:introduction_label}

Micromagnetic simulation connects material parameters, sample geometry, and applied stimuli to magnetic-domain structure and dynamics, supporting the design of magnetic materials and devices \cite{brown1959micromagnetics,abert2019micromagnetics}. The Landau--Lifshitz--Gilbert (LLG) equation \cite{landau1935dispersion,gilbert2004phenomenological} describes magnetization dynamics under the effective field. Repeated field evaluations during relaxation and time integration determine much of the computational cost. Exchange and anisotropy involve local or short-range interactions \cite{abert2019micromagnetics}, whereas the demagnetization field requires an open-boundary magnetostatic calculation that couples the entire sample \cite{newell1993demagnetizing,fredkin1990hybrid}. On regular grids, translational invariance allows the discrete magnetostatic interaction to be evaluated as a convolution using fast Fourier transforms. This approach underlies OOMMF and GPU frameworks including MuMax3, magnum.np, MicroMagnetic.jl, and mumax+ \cite{donahue1999oommf,vansteenkiste2014mumax3,bruckner2023magnumnp,wang2024micromagneticjl,moreels2026mumaxplus}. These methods are computationally efficient for geometries well represented by regular grids. Curved boundaries, irregular geometries, and localized refinement motivate the use of unstructured meshes \cite{abert2013magnumfe,yang2019adaptive_demag}.

Tetrahedral finite element methods accommodate these geometries and permit local refinement, with the open-boundary magnetostatic problem commonly formulated in terms of a scalar potential \cite{fredkin1990hybrid,abert2013magnumfe,abert2019micromagnetics}. Finite element micromagnetic frameworks such as Magpar, Nmag, and magnum.fe provide established tools for simulations on unstructured meshes \cite{scholz2003magpar,fischbacher2007nmag,abert2013magnumfe}. AFEMAG is an adaptive finite element micromagnetics code built on AFEPack \cite{cai2024afepack}. It follows the tetrahedral framework of Yang et al., combining an implicit midpoint discretization of the LLG equation with a PDE-based demagnetization field calculation \cite{yang2021framework}.

Accelerating this finite element workflow requires coordinating irregular mesh operations, dense boundary interactions, and repeated sparse linear solves. Their parallelism and memory access differ from those of regular-grid stencils and FFTs, and their performance depends on indirect data access, sparse operator structure, and host--device transfers \cite{bell2009spmv,sulyok2019locality}. Earlier GPU-accelerated finite element micromagnetics demonstrated speedups whose magnitude depended on the placement of magnetostatic and time-integration tasks \cite{kakay2010gpu_fem}. Reducing the overall simulation time requires coordinating boundary evaluation and linear solves throughout the time integration procedure. The implicit prediction and correction procedure repeatedly couples these operations through field updates and convergence checks. Their data dependencies and reuse must guide the CPU--GPU partition.

In this work, we develop a heterogeneous execution framework for AFEMAG that combines CPU and GPU computation within a single workstation. The CPU performs finite element assembly, controls prediction and correction, checks convergence, and handles output, while the GPU evaluates the dense boundary interaction and solves the sparse systems. Boundary geometry and sparse operators that depend only on the mesh are prepared once. Device data and solver contexts are retained across field evaluations, and vectors updated during iteration are transferred at explicit synchronization points. The boundary stage is implemented in CUDA and the sparse systems are solved with AMGX, with AFEMAG controlling their coupling and data lifetimes. We validate the framework using cases from all five \mumag{} standard problems, covering quasistatic hysteresis, equilibrium configurations, critical sizes, and dynamics driven by applied fields and electric currents \cite{mumag,nist_mumag_activity,donahue1999oommf}. Together, these cases exercise different field terms, convergence criteria, and driving mechanisms within the same heterogeneous implementation. Timings of individual components and complete time steps compare this framework with the original CPU implementation using the same meshes, time steps, and stopping tolerances. The timing analysis distinguishes initial setup from repeated computation and includes data transfers during time stepping.

The remainder of this paper is organized as follows. Section~\ref{sec:numerical_method_label} presents the governing equations and finite element formulation. Section~\ref{sec:gpu_implementation_label} describes the heterogeneous decomposition, data lifecycle, and execution algorithm. Section~\ref{sec:results_label} reports the numerical and performance results, and Section~\ref{sec:conclusion_label} concludes the paper.

\section{Numerical Method}
\label{sec:numerical_method_label}

We use the tetrahedral finite element framework of Yang et al.~\cite{yang2021framework}, with continuous piecewise-linear elements in space, an implicit midpoint rule in time, predictor--corrector iteration, and a PDE-based demagnetization field calculation. The heterogeneous implementation changes how the dominant operations are executed and how their data are managed, but leaves this discretization unchanged.

\subsection{Governing Equation}

Let $\Omega \subset \mathbb{R}^3$ denote the ferromagnetic body and let $\partial \Omega$ be its boundary. The magnetization is denoted by $\Mvec(\mathbf{x},t)$, with constant saturation magnetization $\Ms$. The normalized magnetization is defined as
\begin{equation}
    \mvec(\mathbf{x},t)
    =
    \frac{\Mvec(\mathbf{x},t)}{\Ms},
    \qquad
    |\mvec(\mathbf{x},t)| = 1.
    \label{eq:normalized_magnetization}
\end{equation}

We use the characteristic length $L$, scale fields by $\Ms$, and scale time by $(\gamma_0\Ms)^{-1}$, where $\gamma_0=\mu_0|\gamma|$. Thus $\mathbf{x}_{\mathrm{phys}}=L\mathbf{x}$, $\mathbf{H}_{\mathrm{phys}}=\Ms\mathbf{h}$, and $t_{\mathrm{phys}}=t/(\gamma_0\Ms)$. The dimensionless LLG equation is~\cite{landau1935dispersion,gilbert2004phenomenological}
\begin{equation}
    \frac{\partial \mvec}{\partial t}
    =
    - \mvec \times \Heff
    +
    \alpha \mvec \times
    \frac{\partial \mvec}{\partial t},
    \qquad
    \mathbf{x} \in \Omega,
    \quad
    t \in (0,T],
    \label{eq:dimensionless_llg}
\end{equation}
where $\alpha$ is the Gilbert damping constant and, from this point onward, $\Heff$ denotes the dimensionless effective field $\mathbf{H}_{\mathrm{eff,phys}}/\Ms$. The initial and boundary conditions are
\begin{equation}
    \left\{
    \begin{aligned}
        \mvec(\mathbf{x},0) &= \mvec_0(\mathbf{x}),
        && \mathbf{x} \in \Omega,
        \\
        \frac{\partial \mvec}{\partial \mathbf{n}} &= \mathbf{0},
        && \mathbf{x} \in \partial \Omega,\quad t \in (0,T].
    \end{aligned}
    \right.
    \label{eq:natural_boundary_condition}
\end{equation}
where $\mathbf{n}$ is the outward unit normal vector on $\partial \Omega$. Dividing the physical energy by $\mu_0\Ms^2L^3$ gives
\begin{equation}
    \begin{aligned}
        e_{\mathrm{LL}}
        ={}&
        \int_{\Omega}
        \left[
        \frac{\Aex}{\mu_0 \Ms^2L^2}
        |\nabla \mvec|^2
        +
        f_{\mathrm{ani}}(\mvec)
        \right.
        \\
        &\left.
        {}-
        \frac{1}{2}
        \mvec \cdot \Hdemag
        -
        \mvec \cdot \Happ
        \right]
        d\mathbf{x}.
    \end{aligned}
    \label{eq:dimensionless_energy}
\end{equation}
where $\Aex$ is the exchange stiffness, $\mu_0$ is the permeability of vacuum, $\Hdemag$ is the dimensionless demagnetization field, and $\Happ$ is the dimensionless applied field.

For uniaxial anisotropy, the dimensionless anisotropy density is
\begin{equation}
    f_{\mathrm{ani}}(\mvec)
    =
    \frac{\Ku}{\mu_0 \Ms^2}
    \left[
    1 -
    \left(
    \mathbf{e}_{\mathrm{a}}
    \cdot
    \mvec
    \right)^2
    \right],
    \label{eq:uniaxial_anisotropy_energy}
\end{equation}
where $\Ku$ is the anisotropy coefficient and $\mathbf{e}_{\mathrm{a}}$ is the unit easy axis. The effective field is obtained from the variational derivative of the energy and is decomposed as
\begin{equation}
    \Heff
    =
    \mathbf{h}_{\mathrm{ex}}
    +
    \mathbf{h}_{\mathrm{ani}}
    +
    \Hdemag
    +
    \Happ.
    \label{eq:effective_field_decomposition}
\end{equation}
The exchange and anisotropy fields are $\mathbf{h}_{\mathrm{ex}}=2\epsilon\Delta\mvec$ and $\mathbf{h}_{\mathrm{ani}}=2q(\mathbf{e}_{\mathrm{a}}\cdot\mvec)\mathbf{e}_{\mathrm{a}}$, with $\epsilon=\Aex/(\mu_0\Ms^2L^2)$ and $q=\Ku/(\mu_0\Ms^2)$. Benchmark-specific terms and parameters are given in Section~\ref{sec:results_label}.

\subsection{Temporal Discretization and Predictor--Corrector Iteration}

Let $0 = t^0 < t^1 < \cdots < t^{N_t} = T$ be a partition of the time interval, with
\begin{equation}
    \Delta t = t^{n+1} - t^n.
\end{equation}
The magnetization at time $t^n$ is denoted by $\mvec^n$, and the midpoint value is denoted by $\mvec^{n+1/2}$.

Time is discretized by the implicit midpoint scheme of Yang et al.~\cite{yang2021framework}. The time derivative at the midpoint is approximated by
\begin{equation}
    \left.
    \frac{\partial \mvec}{\partial t}
    \right|_{t=t^{n+1/2}}
    \approx
    \frac{\mvec^{n+1} - \mvec^n}{\Delta t}.
    \label{eq:midpoint_time_derivative}
\end{equation}
This gives the midpoint-discrete LLG equation
\begin{equation}
    \mvec^{n+1}
    =
    \mvec^n
    -
    \Delta t
    \mvec^{n+1/2}
    \times
    \Heff^{n+1/2}
    +
    \alpha
    \mvec^{n+1/2}
    \times
    \left(
    \mvec^{n+1}
    -
    \mvec^n
    \right).
    \label{eq:implicit_midpoint_llg}
\end{equation}

Because $\mvec^{n+1}$ is unknown, a forward-Euler prediction is corrected through the implicit midpoint finite element problem. Let $\widetilde{\mvec}^{(k)}$ be the $k$th predictor--corrector approximation of $\mvec^{n+1}$. The midpoint state in correction $k$ is
\begin{equation}
    \mvec^{n+1/2,(k)}
    = \frac{1}{3}(
    \widetilde{\mvec}^{(k)}
    +
    3\mvec^n
    -
    \mvec^{n-1/2}).
    \label{eq:pc_midpoint}
\end{equation}
The effective field is evaluated at $\mvec^{n+1/2,(k)}$, and $\widetilde{\mvec}^{(k+1)}$ is obtained from the discrete weak form. This correction is repeated until

\begin{equation}
    \left\|
    \widetilde{\mvec}^{(k+1)}
    -
    \widetilde{\mvec}^{(k)}
    \right\|_{L^2(\Omega)}
    <
    \mathrm{tol}_{\mathrm{pc}},
    \label{eq:pc_tolerance}
\end{equation}
or until the number of corrections reaches $n_{\mathrm{pc}}^{\max}$.

After each time step, the nodal magnetization can be renormalized to enforce the unit length constraint:
\begin{equation}
    \mvec_i^{n+1}
    \leftarrow
    \frac{\mvec_i^{n+1}}
    {|\mvec_i^{n+1}|},
    \qquad
    i = 1,\ldots,N_h.
    \label{eq:nodal_renormalization}
\end{equation}
This renormalization limits drift from the unit-length constraint in long simulations.

\subsection{Spatial Discretization on Tetrahedral Meshes}

The computational domain $\Omega$ is discretized by a tetrahedral mesh
\begin{equation}
    \mathcal{T}_h
    =
    \{K_1,K_2,\ldots,K_{N_e}\},
\end{equation}
where $K_j$ denotes a tetrahedral element and $N_e$ is the number of tetrahedra. A continuous piecewise linear finite element space is defined as
\begin{equation}
    V_h
    =
    \left\{
    v_h \in C^0(\Omega)
    :
    v_h|_K \in P_1(K),
    \quad
    \forall K \in \mathcal{T}_h
    \right\}.
    \label{eq:linear_fe_space}
\end{equation}
The discrete magnetization is approximated in $[V_h]^3$:
\begin{equation}
    \mvec_h(\mathbf{x},t)
    =
    \sum_{i=1}^{N_h}
    \mvec_i(t)
    \phi_i(\mathbf{x}),
    \label{eq:fe_magnetization_expansion}
\end{equation}
where $\phi_i$ are the nodal basis functions and $N_h$ is the number of mesh nodes.

Testing the midpoint-discrete LLG equation with $\boldsymbol{\psi}_h \in [V_h]^3$ and integrating over the domain gives the weak form. At each correction step, find $\mvec_h^{n+1} \in [V_h]^3$ such that
\begin{align}
    \int_{\Omega}
    \mvec_h^{n+1}
    \cdot
    \boldsymbol{\psi}_h
    \, d\mathbf{x}
    &=
    \int_{\Omega}
    \mvec_h^n
    \cdot
    \boldsymbol{\psi}_h
    \, d\mathbf{x}
    \nonumber
    \\
    &\quad
    -
    \Delta t
    \int_{\Omega}
    \left(
    \mvec_h^{n+1/2}
    \times
    \mathbf{H}_{\mathrm{eff},h}^{n+1/2}
    \right)
    \cdot
    \boldsymbol{\psi}_h
    \, d\mathbf{x}
    \nonumber
    \\
    &\quad
    +
    \alpha
    \int_{\Omega}
    \left[
    \mvec_h^{n+1/2}
    \times
    \left(
    \mvec_h^{n+1}
    -
    \mvec_h^n
    \right)
    \right]
    \cdot
    \boldsymbol{\psi}_h
    \, d\mathbf{x}.
    \label{eq:weak_form_general}
\end{align}
The homogeneous Neumann boundary condition for the exchange term eliminates the surface contribution after integration by parts.

\subsection{Discrete Demagnetization Field and Gradient Recovery}

The dimensionless demagnetization field is $\Hdemag=-\nabla\phi$, with $\phi=\phi_1+\phi_2$. The first potential satisfies $\Delta\phi_1=\nabla\cdot\mvec$ in $\Omega$ and $\phi_1=0$ on $\partial\Omega$; the second is harmonic in $\Omega$ and carries the open-boundary correction \cite{fredkin1990hybrid,yang2021framework}. For nodal basis functions $\{\phi_i\}_{i=1}^{N_h}$, define
\begin{equation}
    K_{i\ell}=\int_{\Omega}\nabla\phi_i\cdot\nabla\phi_\ell\,d\mathbf{x},
    \qquad
    f_i(\mvec_h)=\int_{\Omega}\mvec_h\cdot\nabla\phi_i\,d\mathbf{x}.
    \label{eq:demag_stiffness_rhs}
\end{equation}
The coefficients of $\phi_{1,h}$ are obtained from the Dirichlet system
\begin{equation}
    K_{II}\boldsymbol{\phi}_{1,I}=\mathbf{f}_I(\mvec_h),
    \qquad
    \boldsymbol{\phi}_{1,B}=\mathbf{0},
    \label{eq:phi1_discrete}
\end{equation}
where $I$ and $B$ denote interior and boundary degrees of freedom. At each boundary target $\mathbf{x}_i$, quadrature gives the Dirichlet data for $\phi_2$ as
\begin{equation}
    b_i
    =
    \sum_{j=1}^{N_q}
    w_j N(\mathbf{x}_i-\mathbf{y}_j)
    \left[
    -\mvec_h(\mathbf{y}_j)\cdot\mathbf{n}_j
    +\nabla\phi_{1,h}(\mathbf{y}_j)\cdot\mathbf{n}_j
    \right],
    \quad
    N(\mathbf{r})=-\frac{1}{4\pi|\mathbf{r}|}.
    \label{eq:discrete_boundary_integral}
\end{equation}
The harmonic correction is then computed from
\begin{equation}
    K_{II}\boldsymbol{\phi}_{2,I}
    =-K_{IB}\mathbf{b},
    \qquad
    \boldsymbol{\phi}_{2,B}=\mathbf{b}.
    \label{eq:phi2_discrete}
\end{equation}
Each field evaluation requires two sparse systems and the dense target--source interaction in Eq.~\eqref{eq:discrete_boundary_integral}.

Following Yang et al.~\cite{yang2021framework}, a superconvergent patch recovery operator $G_h$ is applied to the piecewise linear potential. The field used in the LLG update is $\mathbf{H}_{\mathrm{d},h}=-G_h(\phi_{1,h}+\phi_{2,h})$.

\section{Heterogeneous Computing Design and Implementation}
\label{sec:gpu_implementation_label}

A fixed tetrahedral mesh separates the simulation data into static and dynamic parts. Geometry, quadrature rules, connectivity, sparse matrices, and solver hierarchies remain unchanged throughout a run. Magnetization, right-hand sides, potentials, and boundary values change at every predictor--corrector iteration. These lifetimes determine which data are initialized once, retained on the device, or transferred between host and device.

\subsection{Work Partitioning and Data Lifecycle}

Unstructured finite element workloads combine irregular assembly with algebraic operations that support extensive parallelism. Their performance depends on indirect access and data movement as well as arithmetic \cite{bell2009spmv,sulyok2019locality}. AFEMAG retains mesh traversal, finite element assembly, predictor--corrector control, convergence tests, and output on the multicore CPU. The GPU evaluates interactions between all boundary targets and quadrature sources and performs the repeated sparse solves.

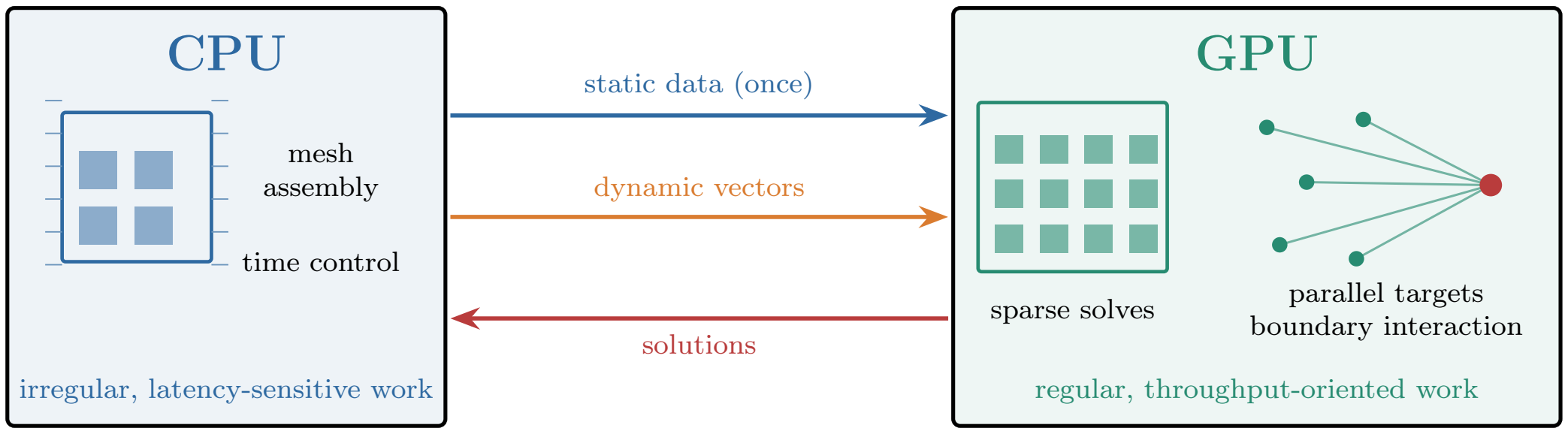
\begin{figure}[htbp]
    \centering
    \resizebox{\textwidth}{!}{%
    \begin{tikzpicture}[
        font=\sffamily,
        >=Stealth,
        flow/.style={-{Stealth},line width=1.2pt},
        chip/.style={draw,line width=1.1pt,rounded corners=2pt},
        labelbox/.style={draw=none,align=center,font=\scriptsize}
    ]
        \definecolor{cpublue}{RGB}{46,105,161}
        \definecolor{gpugreen}{RGB}{39,139,113}
        \definecolor{signalorange}{RGB}{218,124,48}
        \definecolor{targetred}{RGB}{185,60,60}

        \node[chip,fill=cpublue!8,minimum width=4.4cm,minimum height=4.2cm] (cpu) at (2.2,2.2) {};
        \node[font=\bfseries\large,text=cpublue] at (2.2,3.85) {CPU};
        \draw[cpublue,line width=1pt,rounded corners=1pt] (0.55,1.75) rectangle (2.05,3.25);
        \foreach \x in {0.72,1.28}{
            \foreach \y in {1.92,2.48}{
                \fill[cpublue!55] (\x,\y) rectangle ++(0.38,0.38);
            }
        }
        \foreach \y in {1.72,2.05,2.38,2.71,3.04,3.37}{
            \draw[cpublue!70] (0.38,\y) -- (0.55,\y);
            \draw[cpublue!70] (2.05,\y) -- (2.22,\y);
        }
        \node[labelbox,text width=1.75cm] at (3.15,2.65) {mesh assembly};
        \node[labelbox,text width=1.75cm] at (3.15,1.75) {time control};
        \node[font=\scriptsize,text=cpublue] at (2.2,0.45) {irregular, latency-sensitive work};

        \node[chip,fill=gpugreen!8,minimum width=6.1cm,minimum height=4.2cm] (gpu) at (12.55,2.2) {};
        \node[font=\bfseries\large,text=gpugreen] at (12.55,3.85) {GPU};
        \draw[gpugreen,line width=1pt,rounded corners=1pt] (9.75,1.65) rectangle (11.65,3.35);
        \foreach \x in {9.92,10.37,10.82,11.27}{
            \foreach \y in {1.84,2.29,2.74}{
                \fill[gpugreen!62] (\x,\y) rectangle ++(0.28,0.28);
            }
        }
        \node[font=\scriptsize] at (10.7,1.25) {sparse solves};

        \coordinate (target) at (14.9,2.52);
        \foreach \p in {(12.65,3.10),(13.05,2.55),(12.78,1.92),(13.62,3.18),(13.55,1.78)} {
            \draw[gpugreen!65,line width=0.65pt] \p -- (target);
            \fill[gpugreen] \p circle (2.2pt);
        }
        \fill[targetred] (target) circle (3.2pt);
        \node[font=\scriptsize,align=center] at (13.85,1.25) {parallel targets\\boundary interaction};
        \node[font=\scriptsize,text=gpugreen] at (12.55,0.45) {regular, throughput-oriented work};

        \draw[flow,cpublue] (4.45,3.22) -- (9.45,3.22)
            node[midway,above,font=\scriptsize,text=cpublue] {static data (once)};
        \draw[flow,signalorange] (4.45,2.20) -- (9.45,2.20)
            node[midway,above,font=\scriptsize,text=signalorange] {dynamic vectors};
        \draw[flow,targetred] (9.45,1.18) -- (4.45,1.18)
            node[midway,below,font=\scriptsize,text=targetred] {solutions};
    \end{tikzpicture}%
    }
    \caption{Task allocation and data flow in the heterogeneous framework. Data determined by the mesh are initialized once, while vectors updated during iteration are transferred between the CPU and GPU at explicit synchronization points. The diagram of targets and sources illustrates the independent evaluation of boundary sums.}
    \label{fig:heterogeneous_dataflow}
\end{figure}

Figure~\ref{fig:heterogeneous_dataflow} shows the resulting dataflow. During setup, the CPU assembles the mass and magnetostatic matrices and extracts boundary targets, quadrature points, weights, normals, and interpolation data. These static arrays are converted to contiguous storage, transferred once, and retained in device allocations. Separate solver contexts are constructed once for the fixed operators. During each correction, only the current right hand sides, initial guesses, and sampled values of the magnetization and potential gradient are uploaded. Boundary values and solutions of the linear systems return to the CPU for the next assembly or convergence test.

The solution of each sparse system is transferred to the CPU for assembly, after which the updated load vector and initial guess are uploaded to the GPU. The reported timings include these transfers.

\subsection{Data Reuse in Field Evaluation and Sparse Solves}

The dense stage implements Eq.~\eqref{eq:discrete_boundary_integral}. Each boundary value $b_i$ depends on all $N_q$ quadrature sources, whereas different target indices are independent. One device thread is assigned to each target; it holds $\mathbf{x}_i$ fixed and accumulates the source contributions in double precision. The threads write disjoint output entries, so no atomic update is required. Other GPU boundary methods based on direct summation also process targets in parallel and accumulate source contributions sequentially \cite{geng2013gpu_boundary}. Direct evaluation of the boundary sums requires $\mathcal{O}(N_bN_q)$ operations, distributed across GPU threads assigned to independent targets.

Boundary coordinates, quadrature coordinates, normals, weights, and interpolation indices are flattened into contiguous arrays during setup and kept in a persistent device cache. At every field evaluation, the CPU interpolates $\mvec_h$ and $\nabla\phi_{1,h}$ at the quadrature points; only these dynamic samples are copied to preallocated device buffers. The device evaluates all $b_i$, and the boundary vector is returned for the harmonic finite element solve. Preallocation removes repeated geometry packing and device allocation from the nonlinear loop.

The sparse operators have a similar lifetime. For a fixed mesh, the mass and magnetostatic Dirichlet matrices have constant coefficients and sparsity patterns. Their data in compressed sparse row format, preconditioner hierarchies, work vectors, and solver handles are created before time stepping and reused throughout prediction and correction. For each solve, only the right hand side and initial guess are updated. The existing solver context is then used to compute the solution, which is transferred to the CPU. The implementation uses AMGX mode \texttt{dDDI}, with matrix and vector values stored in double precision and indices stored as 32 bit integers \cite{naumov2015amgx}. AFEMAG controls operator construction, solver lifetime, transfer placement, and coupling to the LLG iteration, while AMGX performs the device algebra.

For the reported runs, the mass and magnetostatic systems use flexible GMRES with at most 30 iterations, restart length 32, and a relative-initial-residual $L^2$ tolerance of $10^{-10}$. The preconditioner is an aggregation AMG V-cycle with \texttt{BLOCK\_JACOBI} smoothing, no pre-sweep, three post-sweeps, relaxation factor 0.75, \texttt{SIZE\_2} aggregation, at most 50 levels, and a minimum coarse-grid size of 32 rows. These parameters and the predictor--corrector limit of 16 iterations are held fixed in each CPU--GPU comparison.

\subsection{Heterogeneous Predictor--Corrector Algorithm}

Algorithm~\ref{alg:heterogeneous_workflow} summarizes the heterogeneous time integrator by grouping related operations into major CPU and GPU stages. Interpolation, boundary imposition, and vector transfers follow the dependencies described above but are omitted from the pseudocode. The two magnetostatic solves, direct boundary evaluation, field recovery, and three magnetization solves remain coupled within each correction.

\begin{algorithm}[H]
\caption{Heterogeneous finite element time stepping}
\label{alg:heterogeneous_workflow}
\begin{algorithmic}[1]
\algrenewcommand{\algorithmiccomment}[1]{\hfill\makebox[3em][r]{\textnormal{#1}}}
\State CPU: assemble fixed finite element operators and boundary data.
\State GPU: cache static arrays and initialize reusable solver contexts.
\Statex \hfill\textit{Device}
\For{$n=0,\ldots,N_t-1$}
    \State Predict $\widetilde{\mvec}^{(0)}$ and set $k=0$. \Comment{CPU}
    \Repeat
        \State Form $\mvec^{n+1/2,(k)}$ and assemble the load vector for $\phi_1$.
        \State Solve for $\phi_1$, evaluate $\mathbf{b}$, and solve for $\phi_2$. \Comment{GPU}
        \State Solve the three mass systems for $\widetilde{\mvec}^{(k+1)}$.
        \State Set $k\leftarrow k+1$ and evaluate the update norm. \Comment{CPU}
    \Until{$\|\widetilde{\mvec}^{(k)}-\widetilde{\mvec}^{(k-1)}\|_{L^2}
    <\mathrm{tol}_{\mathrm{pc}}$ or $k=n_{\mathrm{pc}}^{\max}$}
    \State Accept and, if required, renormalize $\mvec^{n+1}$.
\EndFor
\end{algorithmic}
\end{algorithm}

The timings for repeated time steps in Section~\ref{sec:results_label} exclude setup and include the dynamic path inside the loop: vector transfers, sparse solves, direct boundary evaluation, host assembly, and convergence control.

\section{Results}
\label{sec:results_label}

We report a CPU--GPU solver-path consistency test and AFEMAG GPU results for cases from all five \mumag{} problems, together with CPU--GPU timings. Problems 1--3 cover quasistatic and equilibrium calculations. The dynamic tests use the 170 degree field case of Problem 4 and the $\xi=0.05$ case of Problem 5.

The selected cases test quasistatic relaxation, energy competition sensitive to mesh resolution, magnetization dynamics under applied fields, spin transfer torque, direct boundary evaluation, and repeated sparse solves \cite{mumag,nist_mumag_activity,hertel2002finite,lepadatu2020,exl2017,reichel2023,kraft2024}.

\subsection{Computational Environment and Timing Protocol}

All timings were obtained on one compute node with two AMD EPYC 7713 processors (128 physical cores in total), 1~TB RAM, and one NVIDIA A100 80~GB PCIe GPU. The software environment was Ubuntu 22.04, NVIDIA driver 560.35.05, CUDA 12.6, AMGX 2.5.0, and GCC 11.4.0. The code was compiled as C++17 with \texttt{-O3 -fopenmp}; CUDA code used \texttt{-O3 -arch=sm\_80}. CPU and GPU versions used the same mesh, time step, stopping tolerances, and OpenMP thread count shown in each table. GPU timings include host--device vector transfers during repeated solves but exclude one-time mesh construction, finite element assembly, boundary-cache construction, and AMGX setup. Each timing entry was obtained from a single run over the stated number of time steps.

\subsection{Consistency of CPU and GPU Solvers}

We tested the sparse linear solvers on a $1\times1\times0.02~\mu\mathrm{m}^3$ Permalloy film at zero applied field, with $\Ms=8.0\times10^5~\mathrm{A/m}$, $\Aex=1.3\times10^{-11}~\mathrm{J/m}$, $\Ku=0$, and $\alpha=0.2$. The initial magnetization points along $-x$ for $y\geq0$ and $+x$ for $y<0$. The CPU and GPU solvers are used separately to solve the mass systems and the systems for the demagnetizing potential within the same time integration procedure. Both calculations use a dimensionless time step of 0.005. The tolerances for convergence to equilibrium, linear system solution, and prediction and correction are identical in both calculations and are set to $10^{-6}$, $10^{-8}$, and $10^{-8}$, respectively.

\begin{figure}[htbp]
    \centering
    \includegraphics[width=0.48\textwidth]{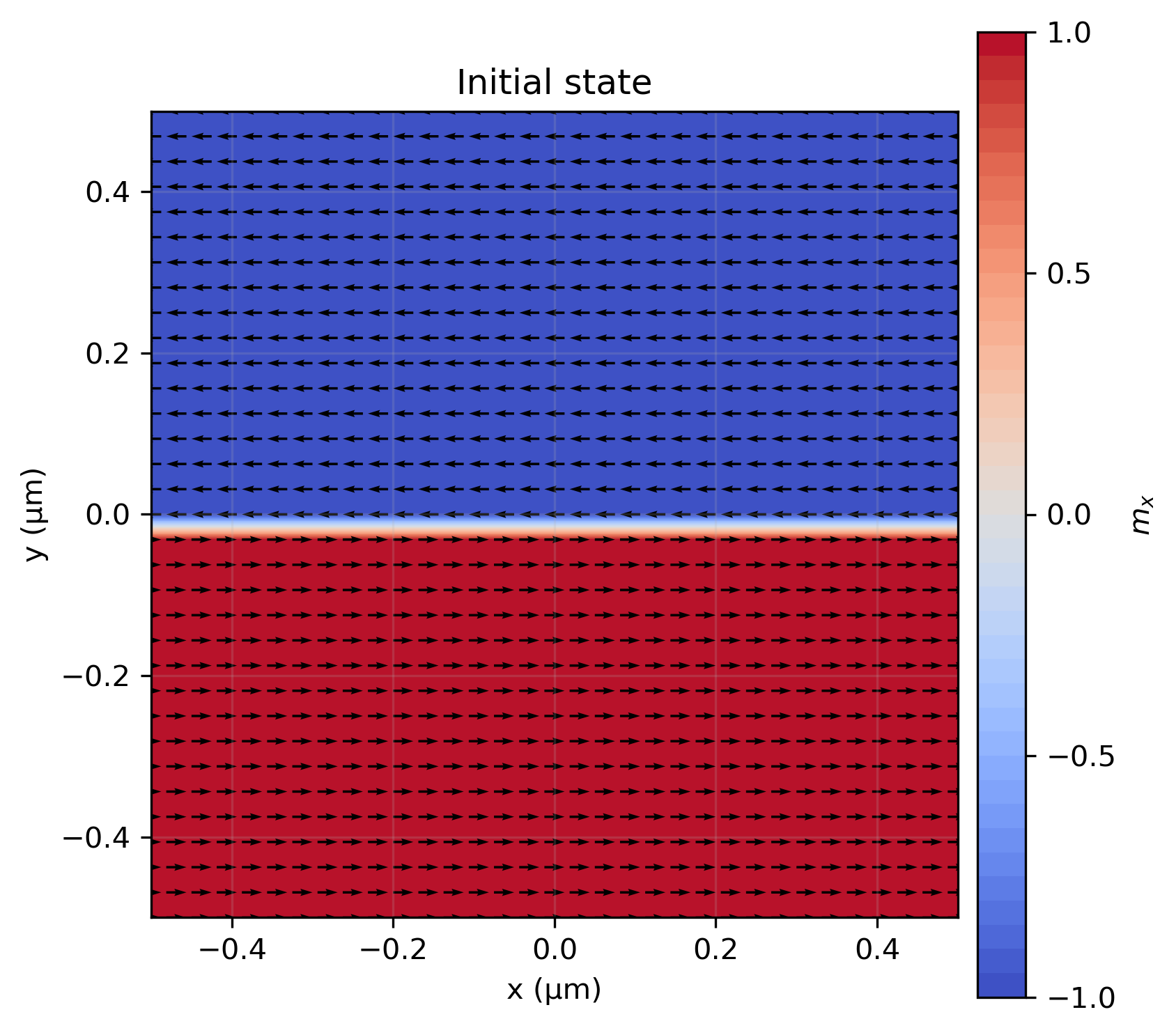}
    \hfill
    \includegraphics[width=0.48\textwidth]{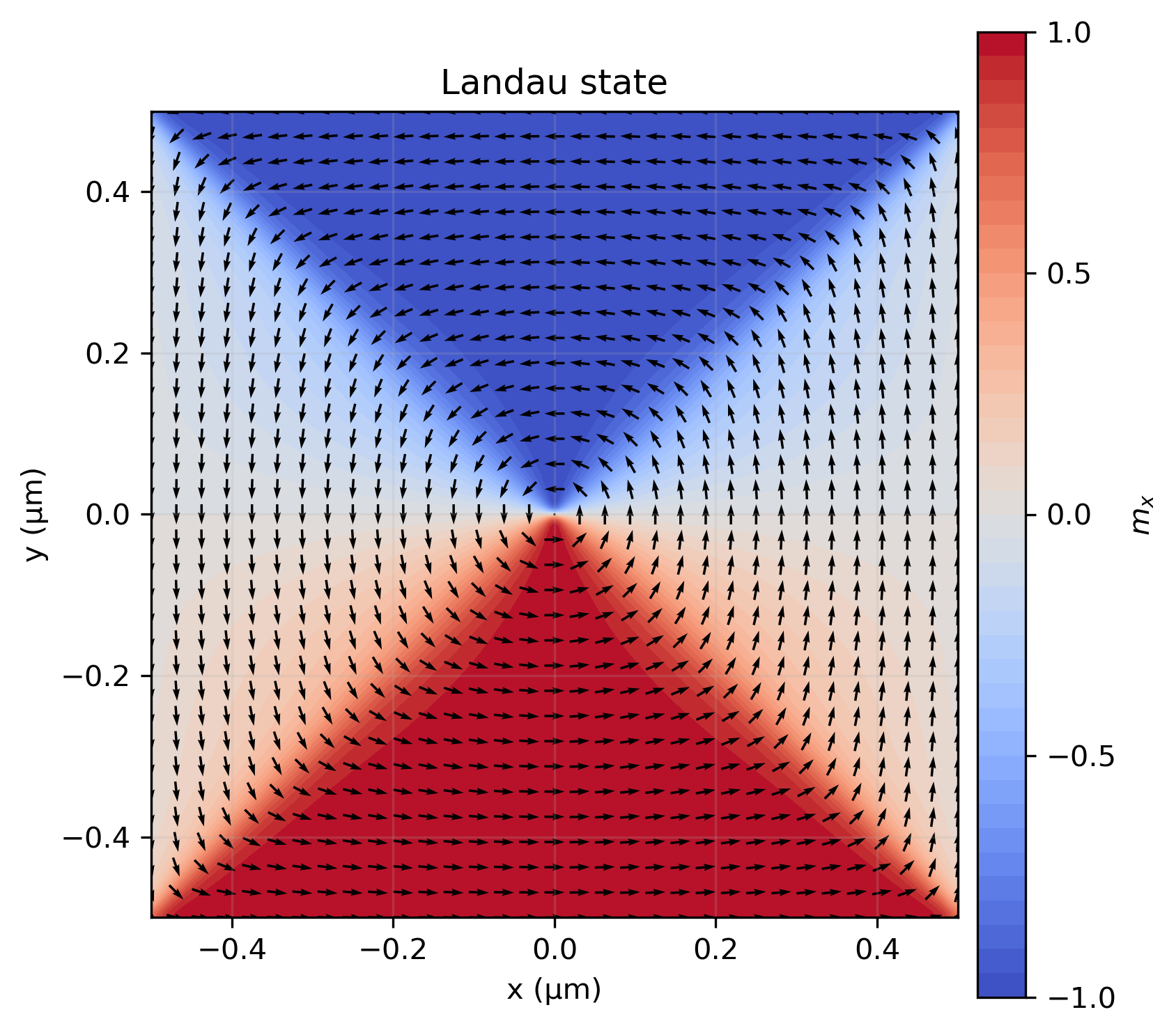}
    \caption{Initial two-domain magnetization (left) and relaxed Landau state on the finest mesh (right).}
    \label{fig:landau_states}
\end{figure}

Figure~\ref{fig:landau_states} shows the initial configuration and the relaxed Landau state on the finest mesh. Table~\ref{tab:landau_energy} compares the total energy density averaged over the sample volume after relaxation on six meshes. The largest relative difference between the two solver paths is $9.23\times10^{-14}$. The small differences reflect the use of different Jacobi-preconditioned conjugate-gradient implementations, with deal.II on the CPU and CUDA on the GPU.

\begin{table}[htbp]
\centering
\caption{Total energy density ($\mathrm{J/m^3}$) at equilibrium computed using CPU and GPU solvers.}
\label{tab:landau_energy}
\small
\begin{tabular}{rcc}
\toprule
Mesh points & CPU & Heterogeneous \\
\midrule
243     & 810.4099656262711  & 810.4099656261963 \\
867     & 1071.6996604379801 & 1071.6996604379790 \\
3,267   & 1279.0982207810328 & 1279.0982207810320 \\
12,675  & 1394.4783908754757 & 1394.4783908754762 \\
66,564  & 1471.4007723955672 & 1471.4007723955683 \\
462,343 & 1493.6825076985840 & 1493.6825076985829 \\
\bottomrule
\end{tabular}
\end{table}

\subsection{Standard Problem 1: Quasistatic Hysteresis}

Standard Problem 1 examines quasistatic hysteresis under an applied field swept along each in-plane axis of a rectangular sample. The geometry and material parameters follow the \mumag{} specification \cite{mumag}. At each field value, the \LLG{} equation is integrated until the relaxation criterion is satisfied, and the relaxed magnetization from the previous field value initializes the next step.

The sample is a $1000\times2000\times20~\mathrm{nm}^3$ Permalloy rectangle with $\Aex=1.3\times10^{-11}~\mathrm{J/m}$, $\Ms=8.0\times10^5~\mathrm{A/m}$, and $\Ku=5.0\times10^2~\mathrm{J/m^3}$. The applied field is swept over $[-50,50]~\mathrm{mT}$ in $2.5~\mathrm{mT}$ increments. The hysteresis calculation uses 15,443 mesh points; performance is measured on 15,443- and 96,051-point meshes.

Table~\ref{tab:sp1_quantitative} reports the coercive field and longitudinal remanence obtained for each direction of the applied field.

\begin{table}[htbp]
\centering
\caption{Computed hysteresis quantities for Standard Problem 1 on the 15,443-point mesh.}
\label{tab:sp1_quantitative}
\begin{tabular}{lcc}
\toprule
Quantity & $H_0 \parallel x$ (short axis) & $H_0 \parallel y$ (long axis) \\
\midrule
Coercive field (mT) & 6.8 & 14.1 \\
Longitudinal remanence ($M_r/M_s$) & 0.55 & 0.89 \\
\bottomrule
\end{tabular}
\end{table}

Both the coercive field and the longitudinal remanence are higher for the field applied along the long axis, consistent with the sample's shape anisotropy.

For each mesh, Table~\ref{tab:sp1_runtime_combined} reports one demagnetization field evaluation and a measured 50-step run with dimensionless step size 0.002.

\begin{table}[htbp]
\centering
\caption{Measured wall time (s) for one demagnetization field evaluation and 50 time steps. Column headings give the OpenMP thread count used by the host.}
\label{tab:sp1_runtime_combined}
\small
\resizebox{\textwidth}{!}{
\begin{tabular}{@{}llcccccc@{}}
\toprule
Points & Version
& \multicolumn{3}{c}{Demag field evaluation}
& \multicolumn{3}{c}{50 time steps} \\
\cmidrule(lr){3-5}
\cmidrule(l){6-8}
& & 8 & 16 & 32 & 8 & 16 & 32 \\
\midrule
15443 & AFEMAG     & 46.99 & 23.63 & 12.75 & 20,683.30 & 11,130.90 & 5,415.50 \\
15443 & AFEMAG GPU & 0.33 & 0.28 & 0.25 & 116.26 & 87.27 & 72.62 \\
96051 & AFEMAG     & 1257.99 & 631.58 & 314.28 & 553,603.90 & 276,798.30 & 138,389.60 \\
96051 & AFEMAG GPU & 1.45 & 1.01 & 0.83 & 1,944.54 & 1,227.82 & 864.38 \\
\bottomrule
\end{tabular}
}
\end{table}

The largest component speedup is $867.6\times$, obtained by comparing the 8-thread CPU boundary evaluation with the corresponding GPU run on the 96,051-point mesh. For 50 complete steps, speedups range from $74.6\times$ to $284.7\times$. The smaller speedups for complete time stepping reflect the additional costs of CPU assembly, convergence checks, and data transfers.

\subsection{Standard Problem 2: Thin Film Hysteresis}

\begin{figure}[htbp]
    \centering

    \begin{subfigure}[b]{0.48\textwidth}
        \centering
        \includegraphics[width=\linewidth]{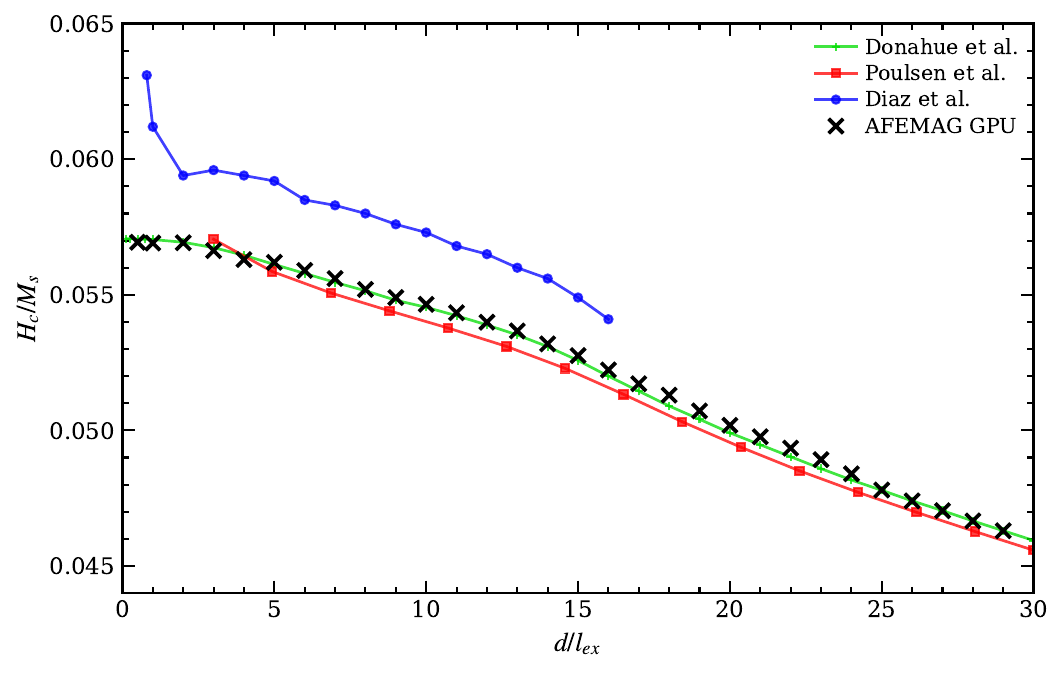}
        \caption{$H_c/M_s$.}
        \label{fig:hc_ms}
    \end{subfigure}
    \hfill
    \begin{subfigure}[b]{0.48\textwidth}
        \centering
        \includegraphics[width=\linewidth]{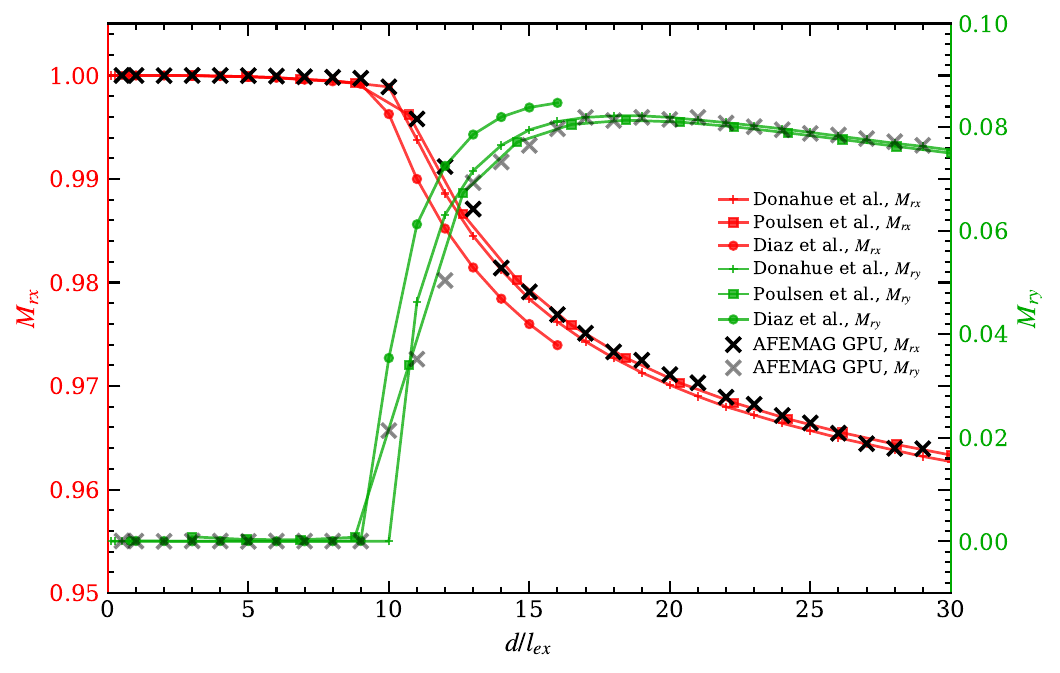}
        \caption{Remanent magnetization.}
        \label{fig:remanence}
    \end{subfigure}

    \caption{Comparison of coercive field and remanent magnetization for Standard Problem 2.}
    \label{fig:sp2_two_panel}
\end{figure}

Standard Problem 2 considers a family of thin films with length $L$, width $d=L/5$, and thickness $t=d/10$ \cite{mumag}. Crystalline anisotropy is neglected. The external field is applied along the $[1,1,1]$ direction and is reduced quasi-statically to obtain the hysteresis response. Following the standard problem definition, the results are reported as functions of the reduced width $d/l_{\mathrm{ex}}$, where
\begin{equation}
    l_{\mathrm{ex}}
    =
    \sqrt{\frac{\Aex}{K_{\mathrm{m}}}},
    \qquad
    K_{\mathrm{m}}
    =
    \frac{\mu_0\Ms^2}{2}.
\end{equation}
Changing $d/l_{\mathrm{ex}}$ alters the balance between exchange and magnetostatic energy while keeping the same scaled thin-film geometry.
Figure~\ref{fig:sp2_two_panel} compares the present results with published reference data for the coercive field and remanent magnetization.

As $d/l_{\mathrm{ex}}$ increases, $H_c/M_s$ decreases from about $5.7\times10^{-2}$ to $4.6\times10^{-2}$. The AFEMAG GPU points follow the reference slope and magnitude, with a small positive deviation in the middle of the interval.

For $d/l_{\mathrm{ex}}<10$, $M_{rx}$ remains near one and $M_{ry}$ near zero. Between approximately 10 and 13, $M_{rx}$ falls while $M_{ry}$ rises, reproducing the remanent-mode transition. The reference datasets show the largest differences in this interval. The computed results reproduce the decrease in $M_{rx}$ and the corresponding increase in $M_{ry}$ across this transition interval.

\subsection{Standard Problem 3: Critical Size Behavior}
\label{sec:sp3_results}

Standard Problem 3 examines the competition between flower and vortex states in a cubic magnetic particle with side length $L$ \cite{mumag,hertel2002finite}. The energies of the two relaxed states are compared to determine the critical side length at which the energetically preferred configuration changes. As in Standard Problem 2, the cube side length is expressed in units of the exchange length
\begin{equation}
    l_{\mathrm{ex}}
    =
    \sqrt{\frac{\Aex}{K_{\mathrm{m}}}},
    \qquad
    K_{\mathrm{m}}
    =
    \frac{\mu_0\Ms^2}{2}.
\end{equation}
A uniaxial anisotropy with magnitude $\Ku = 0.1K_{\mathrm{m}}$ is applied along a principal axis of the cube. In the finite element calculation, the computational domain is represented by a unit cube and different physical side lengths are obtained through length rescaling. For each value of $L$, the system is relaxed from the flower and vortex initial conditions, and the total energies of the two relaxed states are compared. The critical side length $L_c$ is determined from
\begin{equation}
    E_{\mathrm{flower}}(L_c)
    =
    E_{\mathrm{vortex}}(L_c).
    \label{eq:sp3_critical_length_condition}
\end{equation}

On the 2,547-, 17,969-, and 137,313-point meshes, the crossings move from 8.1744 to 8.3300 and 8.4081. The successive increments are approximately halved. Assuming that the observed ratio between successive increments persists, Richardson extrapolation gives an estimated critical length of $L_c\approx8.4869$. A separate 1,073,345-point calculation crosses near 8.4475, inside the same transition interval but about 0.04 below the extrapolated value.

\begin{figure}[htbp]
    \centering

    \begin{subfigure}[b]{0.48\textwidth}
        \centering
        \includegraphics[width=\linewidth]{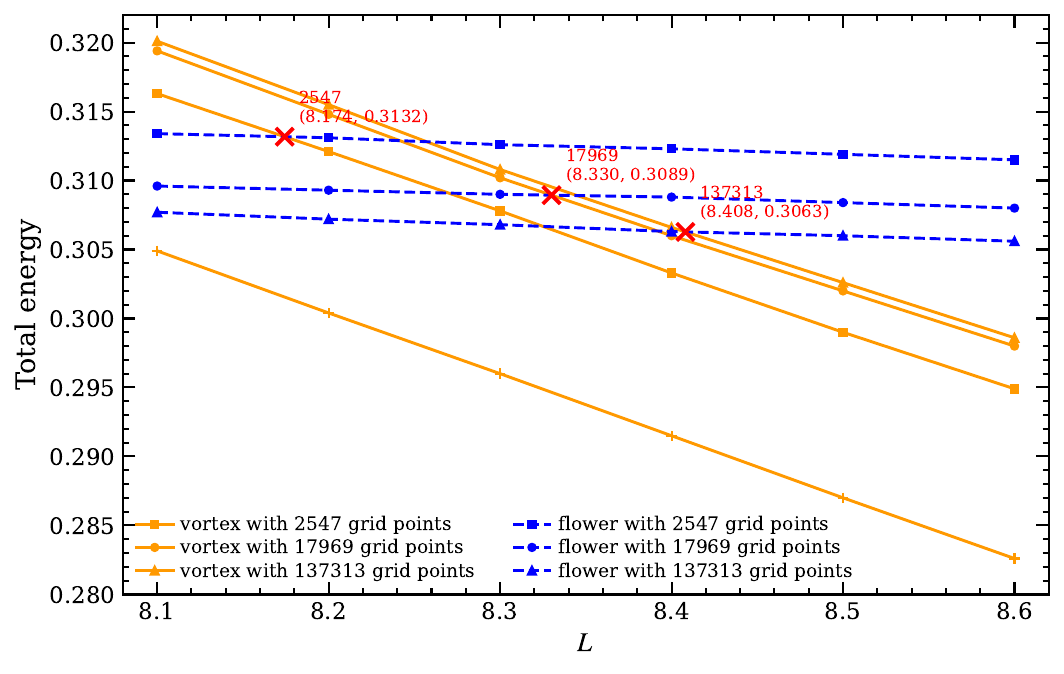}
        \label{fig:std3_energy}
    \end{subfigure}
    \hfill
    \begin{subfigure}[b]{0.48\textwidth}
        \centering
        \includegraphics[width=\linewidth]{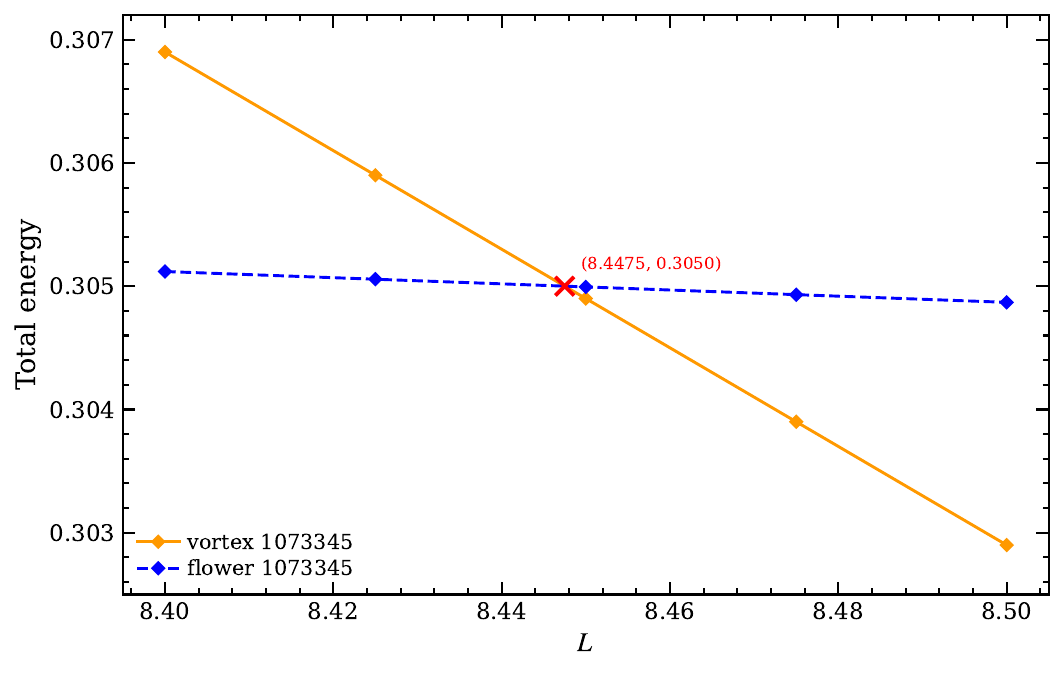}
        \label{fig:std3_energy_1073345}
    \end{subfigure}
    \caption{Flower- and vortex-state energies in Standard Problem 3. Left: three mesh refinements used for the crossing sequence. Right: direct crossing on the 1,073,345-point mesh.}
    \label{fig:sp3_two_panel}
\end{figure}

Table~\ref{tab:sp3_l_comparison} gives the quantitative comparison for the critical side length.

\begin{table}[htbp]
\centering
\caption{Comparison of critical length $L$ for $\mu$MAG Standard Problem 3.}
\label{tab:sp3_l_comparison}
\resizebox{\textwidth}{!}{
\begin{tabular}{lcccccccc}
\toprule
Source 
& This work 
& Ref\cite{yang2021framework}
& Ref\cite{bjork2021}
& Ref\cite{schaffer2024constraint} 
& Ref\cite{nonaka2026magnex}
& Ref\cite{hertel2002finite}
& Ref\cite{selke2013design}
& MuMax3~\cite{vansteenkiste2014mumax3} \\
\midrule
$L$ 
& 8.4869 
& 8.4884 
& 8.477$\pm$0.007 
& 8.4709 
& 8.4708 
& 8.5200 
& 8.4700 
& 8.4700 \\
\bottomrule
\end{tabular}
}
\end{table}

\begin{figure}[htbp]
    \centering

    \begin{subfigure}[b]{0.48\textwidth}
        \centering
        \includegraphics[width=\linewidth]{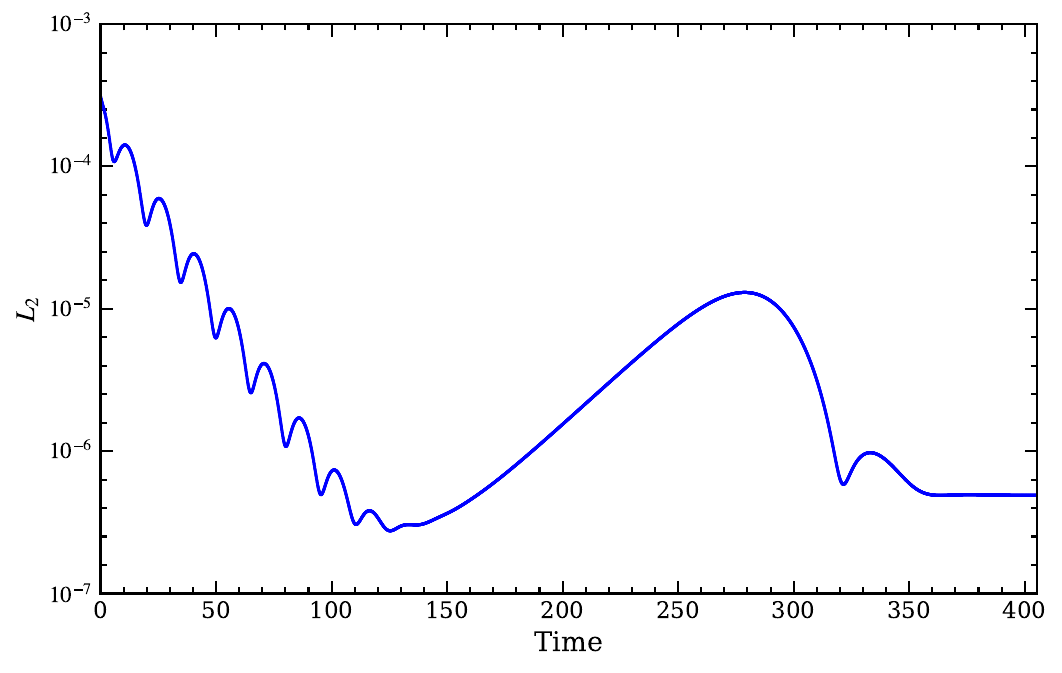}
        \label{fig:l2_vs_time}
    \end{subfigure}
    \hfill
    \begin{subfigure}[b]{0.48\textwidth}
        \centering
        \includegraphics[width=\linewidth]{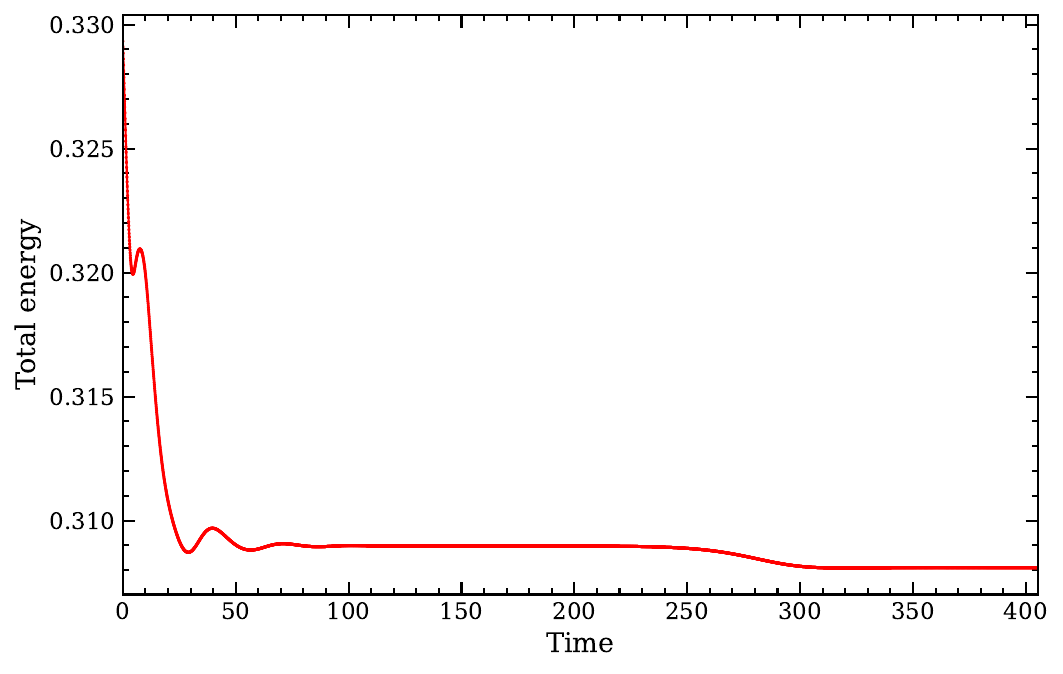}
        \label{fig:total_energy_vs_time}
    \end{subfigure}

    \caption{Relaxation of the twisted flower state on a mesh with 17,969 nodes: $L^2$ norm of the difference between successive states (left) and total energy (right).}
    \label{fig:convergence_history}
\end{figure}

Figure~\ref{fig:convergence_history} shows the relaxation history for the twisted flower state on the mesh with 17969 grid points. The $L^2$ norm of the difference between successive states decreases by several orders of magnitude, while the total energy approaches a plateau.

Table~\ref{tab:sp3_100_steps_runtime} compares the time required for two versions of AFEMAG to run 100 time steps on different meshes under the same time step and tolerance.
\begin{table}[htbp]
\centering
\caption{Runtime for advancing 100 time steps in Standard Problem 3.}
\label{tab:sp3_100_steps_runtime}
\small
\resizebox{\textwidth}{!}{
\begin{tabular}{@{}lcccc@{}}
\toprule
Mesh
& \multicolumn{2}{c}{16 OMP threads}
& \multicolumn{2}{c}{32 OMP threads} \\
\cmidrule(lr){2-3}
\cmidrule(l){4-5}
& AFEMAG original & AFEMAG GPU
& AFEMAG original & AFEMAG GPU \\
\midrule
17969  & 1018.3  & 249.3  & 599.7   & 175.2 \\
137313 & 23136.7 & 1216.0 & 14011.7 & 818.3 \\
\bottomrule
\end{tabular}
}
\end{table}
Table~\ref{tab:sp3_100_steps_runtime} gives end-to-end speedups of $4.1\times$ and $3.4\times$ on the 17,969-point mesh at 16 and 32 OpenMP threads, respectively. On the 137,313-point mesh, the corresponding speedups are $19.0\times$ and $17.1\times$; at 32 threads, the runtime falls from $14,011.7~\mathrm{s}$ to $818.3~\mathrm{s}$.

The finer mesh gives a larger speedup at both thread counts. More CPU threads reduce the runtime of both implementations, while the heterogeneous version remains faster at each thread count. During relaxation in Standard Problem 3, field evaluation and sparse linear solves are repeated using the workflow described in Section~\ref{sec:gpu_implementation_label}.

\subsection{Standard Problem 4: Field Driven Dynamics}

Standard Problem 4 validates time integration and evaluation of the effective field through magnetization dynamics driven by an applied field. Small differences in the demagnetization field, time integration scheme, or numerical damping can shift the averaged magnetization trajectory \cite{lepadatu2020,exl2017,reichel2023,kraft2024}. Following the \mumag{} specification, the benchmark starts from an equilibrium S state obtained by saturating the sample along the $[1,1,1]$ direction and slowly reducing that field to zero. A reversing field is then applied instantaneously at $t=0$, and the subsequent relaxation toward the new field direction is measured through the spatially averaged magnetization.

The simulated sample is a Permalloy thin film of size $500\times125\times3~\mathrm{nm}^3$, with $\Aex=1.3\times10^{-11}~\mathrm{J/m}$, $\Ms=8.0\times10^5~\mathrm{A/m}$, $\Ku=0$, and Gilbert damping $\alpha=0.02$. In the comparison shown here, the applied field is the standard 170 degree case,
$\mu_0\mathbf{H}=[-24.6,4.3,0.0]~\mathrm{mT}$.
The finite element calculation uses a tetrahedral mesh with 835,725 grid points. The reported observables are the normalized volume-averaged components $\langle m_x \rangle$, $\langle m_y \rangle$, and $\langle m_z \rangle$. The curves labeled Ref are taken from \cite{bjork2021}; the MuMax comparison uses \cite{vansteenkiste2014mumax3}.

\begin{figure}[htbp]
    \centering
    \includegraphics[width=0.92\textwidth]{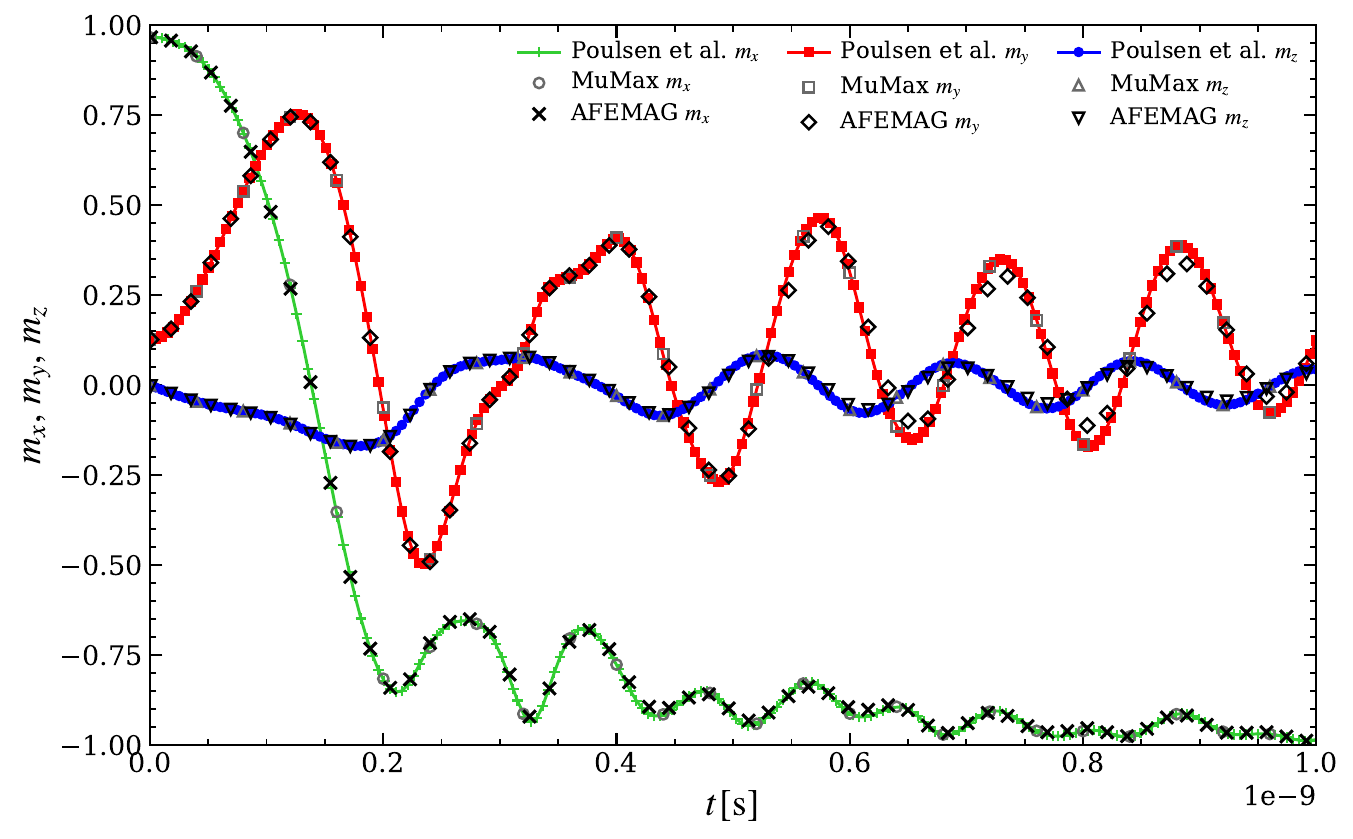}
    \caption{Volume-averaged magnetization for the 170-degree field case of Standard Problem 4, using 835,725 mesh points.}
    \label{fig:sp4_magnetization_time}
\end{figure}

Figure~\ref{fig:sp4_magnetization_time} shows the time dependent averaged magnetization components.

The AFEMAG GPU trajectory follows the reference and MuMax curves over 1~ns. The component $\langle m_x\rangle$ crosses zero near 0.14~ns, and the first $\langle m_y\rangle$ peak occurs near 0.12~ns. The plotted trajectories also show similar oscillation phases and damping over the simulated interval.

\subsection{Standard Problem 5: Current Driven Dynamics}

Standard Problem 5 tests micromagnetic solvers by examining magnetization dynamics driven by an electric current through spin transfer torque \cite{mumag,nist_mumag_activity}. Compared with Problem 4, it adds transport-induced terms to the magnetization dynamics. The physical observable is the motion of a magnetic vortex in a square film under an in-plane current.

Following the \mumag{} specification, the sample is a $100\times100\times10~\mathrm{nm}^3$ Permalloy film centered at the origin and aligned with the Cartesian axes. The material parameters are $\Ms=8.0\times10^5~\mathrm{A/m}$, $\Aex=1.3\times10^{-11}~\mathrm{J/m}$, $\Ku=0$, $\alpha=0.1$, and $\gamma_0=2.211\times10^5~\mathrm{m/(A\,s)}$. The initial state is first constructed as a vortex-like magnetization field,
$\mathbf{f}(x,y,z)=(-y,x,R)$ with $R=10~\mathrm{nm}$, normalized to $\mathbf{m}=\mathbf{f}/|\mathbf{f}|$, and then relaxed without current to obtain the equilibrium vortex used as the initial condition for the current driven simulations.

\begin{figure}[H]
    \centering
    \includegraphics[width=0.62\textwidth]{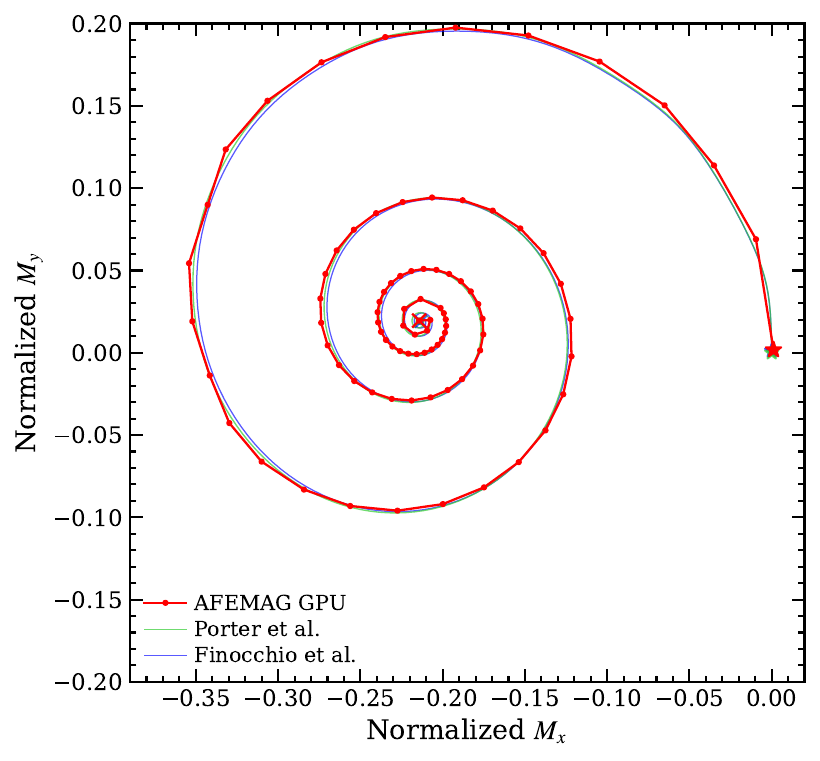}
    \caption{Normalized spatial averages $M_y/M_s$ versus $M_x/M_s$ for the $\xi=0.05$ case of Standard Problem 5.}
    \label{fig:sp5_normalized_my_vs_mx}
\end{figure}

The applied current is in the positive $x$ direction, with $PJ=10^{12}\hat{\mathbf{x}}~\mathrm{A/m^2}$. In physical coordinates, the form implemented for this benchmark is
\begin{equation}
    \frac{\partial \mvec}{\partial t}
    =
    -
    \gamma_0 \mvec \times \mathbf{H}_{\mathrm{eff}}
    +
    \alpha
    \mvec
    \times
    \frac{\partial \mvec}{\partial t}
    -u\frac{\partial \mvec}{\partial x}
    +
    \xi u\left(\mvec\times\frac{\partial \mvec}{\partial x}\right),
    \label{eq:sp5_llg_with_torque}
\end{equation}
where $u$ is the current induced velocity parameter and $\xi$ is the degree of non-adiabaticity. The case shown in Figure~\ref{fig:sp5_normalized_my_vs_mx} uses $\xi=0.05$, corresponding to $u=-72.17~\mathrm{m/s}$ in the standard problem definition. As recommended by the benchmark, the reported quantities are the normalized spatial averages $M_x/M_s$ and $M_y/M_s$, which serve as a proxy for the vortex position as long as the single-vortex structure is preserved.

For $\xi=0.05$, the AFEMAG GPU trajectory reproduces the clockwise spiral, the extent of the outer orbit, and the final displacement seen in the reference curves. Differences are most pronounced in the inner turns.

\section{Conclusion}
\label{sec:conclusion_label}

AFEMAG was reorganized into a heterogeneous execution framework combining CPU and GPU computation while retaining its tetrahedral discretization and implicit midpoint scheme. Static mesh data are separated from vectors updated during iteration, and boundary caches and contexts for sparse solvers are reused throughout prediction and correction. Across six meshes, the CPU and GPU solvers produce energy densities for the relaxed Landau state with a maximum relative difference of $9.23\times10^{-14}$. Cases from all five \mumag{} standard problems test static configurations and magnetization dynamics driven by applied fields and electric currents. For Standard Problem 3 on a mesh with 137,313 nodes, the runtime for 100 time steps decreases from 14,011.7~s using 32 CPU threads to 818.3~s using the same CPU configuration and one NVIDIA A100 GPU, giving a $17.1\times$ speedup.



\bibliographystyle{elsarticle-num}
\bibliography{references}

\appendix


\end{document}